\documentclass{aa}
\usepackage{graphicx}
\usepackage{txfonts}
\newcommand{\etal}{et\,al.\ }
\newcommand{\logg}{\mbox{$\log g$}}
\newcommand{\Teff}{\mbox{$T_\mathrm{eff}$}}
\newcommand{\hh}{\object{H1504$+$65}}
\newcommand{\rxj}{\object{RX\,J2117.1$+$3412}}
\newcommand{\rxja}{\object{RX\,J0122.9$-$7521}}
\newcommand{\elv}{\object{PG1144$+$005}}
\newcommand{\elf}{\object{PG1159$-$035}}

\newcommand{\vier}{\object{PG1424$-$535}}
\newcommand{\fuenf}{\object{PG1520$+$525}}
\newcommand{\sieben}{\object{PG1707$+$427}}
\newcommand{\mct}{\object{MCT0130$-$1937}}
\newcommand{\ngc}{\object{NGC\,246}}

\newcommand{\keins}{\object{K1$-$16}}

\newcommand{\lppr}{\stackrel{<}{\scriptstyle \sim}}

\newcommand{\Htwo}{\mbox{H$_{2}$}}
\begin{document}
   \title
   {Identification of neon in FUSE and VLT spectra of extremely hot 
hydrogen-deficient (pre-) white dwarfs\thanks
{Based on observations made with the NASA-CNES-CSA Far Ultraviolet 
Spectroscopic Explorer. FUSE is operated for NASA by the Johns Hopkins
University under NASA contract NAS5-32985.}\ \thanks
{Based on observations at the Paranal Observatory of the European Southern
   Observatory for programs No.\ 165.H-0588(A), 167.D-0407(A), and 69.D-0719(A).}
   }
 
   \author{K. Werner$^1$, T. Rauch$^{1,2}$, E. Reiff$^1$, J.W. Kruk$^3$ \and R. Napiwotzki$^4$}
   \offprints{K\@. Werner}
   \mail{werner@astro.uni-tuebingen.de}
 
   \institute
    {
     Institut f\"ur Astronomie und Astrophysik, Universit\"at T\"ubingen, Sand 1, D-72076 T\"ubingen, Germany
\and
Dr.-Remeis-Sternwarte, Universit\"at Erlangen-N\"urnberg, Sternwartstra\ss e 7, D-96049 Bamberg, Germany
\and
Department of Physics and Astronomy, Johns Hopkins University,
Baltimore, MD 21218, U.S.A.
\and
Department of Physics and Astronomy, University of Leicester, University
    Road,  Leicester, LE1 7RH, UK
}
    \date{Received xxx / Accepted xxx}
   \authorrunning{K. Werner et al.}
   \titlerunning{Neon in FUSE and VLT spectra of extremely hot H-deficient (pre-) white dwarfs}
   \abstract{One of the strongest absorption lines observed in far-ultraviolet
FUSE spectra of many PG1159 stars remained unidentified up to now. We show that
this line, located at 973.3\AA, stems from \ion{Ne}{vii}. We also present new
optical high-resolution spectra of PG1159 stars, obtained with the ESO VLT,
which cover the \ion{Ne}{vii}~3644\AA\ line and a newly identified \ion{Ne}{vii}
multiplet in the 3850--3910\AA\ region. 
We compare the observed neon lines with NLTE models and conclude a
  substantial neon overabundance in a number of objects. Although a detailed
  analysis is still to be performed in order to compare quantitatively the
  abundances with evolutionary theory predictions, 
this
corroborates the idea that the PG1159 stars and their immediate progenitors, the
[WC]-type nuclei of planetary nebulae, display intershell matter of their
precursor AGB stars. Possibly as the consequence of a late He-shell flash, H-deficient
and (s-processed) Fe-depleted matter, that is strongly enriched by
3$\alpha$-processed elements (C, O, Ne), is dredged up to the surface. Hence, a
detailed study of the element abundance patterns in these peculiar stars gives
the unique possibility to probe mixing and nucleosynthesis processes in
the precursor AGB stars.
        }
   \maketitle

\section{Introduction} 

The PG1159 stars are hot hydrogen-deficient (pre-) white dwarfs ($T_{\rm
eff}$=75\,000--200\,000\,K, $\log g$=5.5--8 [cgs]; Werner 2001). They are
probably the outcome of a late helium-shell flash, a phenomenon that drives the
currently observed fast evolutionary rates of three well-known objects (FG~Sge,
Sakurai's object, V605 Aql). Flash-induced envelope mixing produces a
H-deficient stellar surface. The photospheric composition then essentially
reflects that of the region between the H-and He-burning shells in the precursor
AGB star. The He-shell flash transforms the star back to an AGB star and the
subsequent, second post-AGB evolution explains the existence of Wolf-Rayet
central stars of planetary nebulae and their successors, the PG1159 stars.

The detection of a \ion{Ne}{vii} line at 3644\AA\ in three PG1159 stars (\rxj,
NGC\,246, \keins; Werner \& Rauch 1994, {\small WR94}) and the derived, high
abundance of Ne=2\% (by mass; i.e.\ 20 times solar)  strengthened the view that
these stars display intershell matter strongly enriched by Ne from 3$\alpha$
burning. The next positive result of a search for neon was obtained from EUVE
spectra of \hh, which showed that very strong and broad \ion{Ne}{vii} and
\ion{O}{vi} lines dominate the soft X-ray region (100--130\AA; Werner \& Wolff
1999). From these data and the \ion{Ne}{vii}~3644\AA\ line in a Keck spectrum, a
neon abundance of 2--5\% was derived. Dozens of lines from
\ion{Ne}{vi}--\ion{Ne}{viii} were later detected in the Chandra spectrum of \hh,
confirming this result (Werner \etal 2004, {\small W04}). This recent analysis
also took advantage of FUSE (Far Ultraviolet Spectroscopic Explorer) spectra, in
which we detected for the very first time in any stellar spectrum another
\ion{Ne}{vii} line, located at 973\AA. It is among the strongest absorption
features in that FUSE spectrum.

In this paper we present the results of our systematic search for this newly
identified FUV line in FUSE spectra of a number of PG1159 stars. We also present
new high-resolution optical spectra obtained with ESO's VLT (Very Large
Telescope),  which allow for a much deeper search for the \ion{Ne}{vii}~3644\AA\
line, compared to our earlier {\small WR94} investigation. We also announce the
first detection of a \ion{Ne}{vii} multiplet at 3850--3910\AA.  It will be shown
that the \ion{Ne}{vii}~973\AA\ line is much stronger than the optical neon
lines, and that it can still be observed in objects with photospheric parameters
for which the optical lines are no longer detectable. In contrast to the optical
lines, the FUV line is even detectable at a solar abundance level and can thus
be used as a unique tool for the  neon abundance determination in very hot
H-rich central stars.

In Sect.\,2 we describe our observations, followed by a presentation of our model
atmospheres in Sect.\,3. In Sect.\,4 we compare observed and computed
\ion{Ne}{vii} line profiles. We discuss implications of our results on interior
processes in the precursor AGB stars in the final Section~5.

\begin{table}
\caption{Atmospheric parameters of the program stars as taken from the
literature. The neon abundance has been set to 2\%. 
Abundances are given in \% mass fraction. \label{objects_tab}}
\begin{tabular}{l r r r r r r r r l l}
      \hline
      \hline
      \noalign{\smallskip}
Object  & \Teff & \logg & H & He & C  & O  & ref.\\
        & [kK]  & (cgs) &   &    &    &    &   \\
      \noalign{\smallskip}
      \hline
      \noalign{\smallskip}

\rxja       & 180 & 7.5 &   & 66 & 21 & 11 & A\\
\rxj        & 170 & 6.0 &   & 38 & 54 &  6 & B\\
PG1520+525  & 150 & 7.5 &   & 43 & 38 & 17 & J\\
PG1144+005  & 150 & 6.5 &   & 38 & 58 &  2 & C\\
NGC\,246    & 150 & 5.7 &   & 62 & 30 &  6 & F\\
PG1159-035  & 140 & 7.0 &   & 33 & 48 & 17 & I,J\\
Abell 21    & 140 & 6.5 &   & 33 & 48 & 17 & F\\
\keins      & 140 & 6.4 &   & 33 & 48 & 17 & G\\   
Longmore 3  & 140 & 6.3 &   & 38 & 54 &  6 & F\\
PG1151-029  & 140 & 6.0 &   & 33 & 48 & 17 & F\\
Longmore 4  & 120 & 5.5 &   & 45 & 42 & 11 & D\\
PG1424+535  & 110 & 7.0 &   & 49 & 43 &  6 & J\\
Abell 43    & 110 & 5.7 & 35& 41 & 21 &  1 & E\\      
NGC\,7094   & 110 & 5.7 & 35& 41 & 21 &  1 & E\\      
Abell 30    & 110 & 5.5 &   & 33 & 48 & 17 & H\\
MCT0130-1937&  90 & 7.5 &   & 73 & 22 &  3 & J\\
PG1707+427  &  85 & 7.5 &   & 43 & 38 & 17 & J\\
 \noalign{\smallskip}
      \hline
     \end{tabular}
References in last column: 
A: Werner \etal 1996a,
B: Werner \etal 1996b,
C: Werner \& Heber 1991,
D: Werner \etal 1992,
E: Dreizler \etal 1997,
F: Rauch \& Werner 1997,
G: Kruk \& Werner 1998,
H: Werner \etal 2003a,
I: Werner \etal 1991,
J: Dreizler \& Heber 1998
\end{table}

\begin{table}
\caption{Log of FUSE and VLT observations of our program stars. VLT observations
  were performed with two different instrumental setups (see text).
\label{spectra_tab} }
\begin{tabular}{l| c c |c |c}
\hline 
\hline 
\noalign{\smallskip}
Object &   FUSE   &                    &VLT/1               & VLT/2\\
       &  dataset & t$_{\mathrm{exp}}$ & t$_{\mathrm{exp}}$ & t$_{\mathrm{exp}}$ \\
       &  name    &[min]               &[min]               &[min] \\
\noalign{\smallskip}
\hline
\noalign{\smallskip}
\noalign{\smallskip}
\rxja       &    --    &  -- & 10 & - \\  
\rxj        & P1320501 & 137 & -- & - \\
PG1520+525  & P1320101 &  81 & -- & - \\
PG1144+005  & P1320201 & 114 & 10 & 60\\
NGC\,246    &    --    &  -- &  5 & - \\
PG1159-035  & Q1090101 & 105 &  5 & 60\\
Abell 21    &    --    &  -- & 10 & - \\
\keins      & M1031001-8 & 665 & -- & - \\
Longmore 3  &    --    &  -- &  5 & - \\
PG1151-029  &    --    &  -- & 10 & - \\
Longmore 4  & B0230201 & 434 & 10 & 60\\ 
PG1424+535  & P1320301 & 185 & -- & - \\
Abell 43    &    --    &  -- &  5 & - \\
NGC\,7094   &    --    &  -- &  5 & - \\
Abell 30    &    --    &  -- &  5 & - \\ 
MCT0130-1937&    --    &  -- & 10 & - \\ 
PG1707+427  & P1320401 & 243 & -- & - \\
\noalign{\smallskip}
\hline
\end{tabular}
\end{table}

\section{FUSE and VLT observations}

Our sample consists of 17 PG1159 stars, which represent about 50\% of all known
objects in this spectroscopic class. Their basic atmospheric parameters are
listed in Table\,\ref{objects_tab}. The sample contains two so-called
hybrid-PG1159 stars, which display hydrogen Balmer lines in their spectra
(Abell~43 and NGC\,7094). Traces of nitrogen (typically 1\% by mass) were
discovered in some PG1159 stars. This is neglected in Table\,\ref{objects_tab},
but will be discussed in the last Section.

In Table\,\ref{spectra_tab} we list the FUSE and VLT spectra available for our
program stars.  The FUSE instrument consists of four coaligned telescopes, each
with a prime-focus spectrograph.    Descriptions of the FUSE instrument, and
channel alignment and wavelength calibration issues are given by Moos \etal
(2000) and Sahnow \etal (2000).  The LWRS spectrograph apertures were used for
all observations other than \keins, which was observed in all 3 apertures.  As a
result, the zero-point of the wavelength scale is uncertain to within about $\pm
0.15$\AA.  All exposures were photometric, or nearly so, in all channels for the
LWRS observations.  Exposures in the HIRS and MDRS aperture \keins\ observations
that had less than 50\% peak flux were discarded; the rest were renormalized to
match the LWRS data.    All data were obtained in time-tag mode except for
\rxj, which was observed in histogram mode, and all were processed
with CALFUSE v2.4.    Spectra from individual exposures were coaligned in
wavelength on a channel by channel basis using narrow interstellar absorption
lines, averaged weighted by exposure time, and then shifted to put the
photospheric \ion{O}{vi} features at zero velocity.

\begin{figure*}[tbp]
\includegraphics[width=17.5cm]{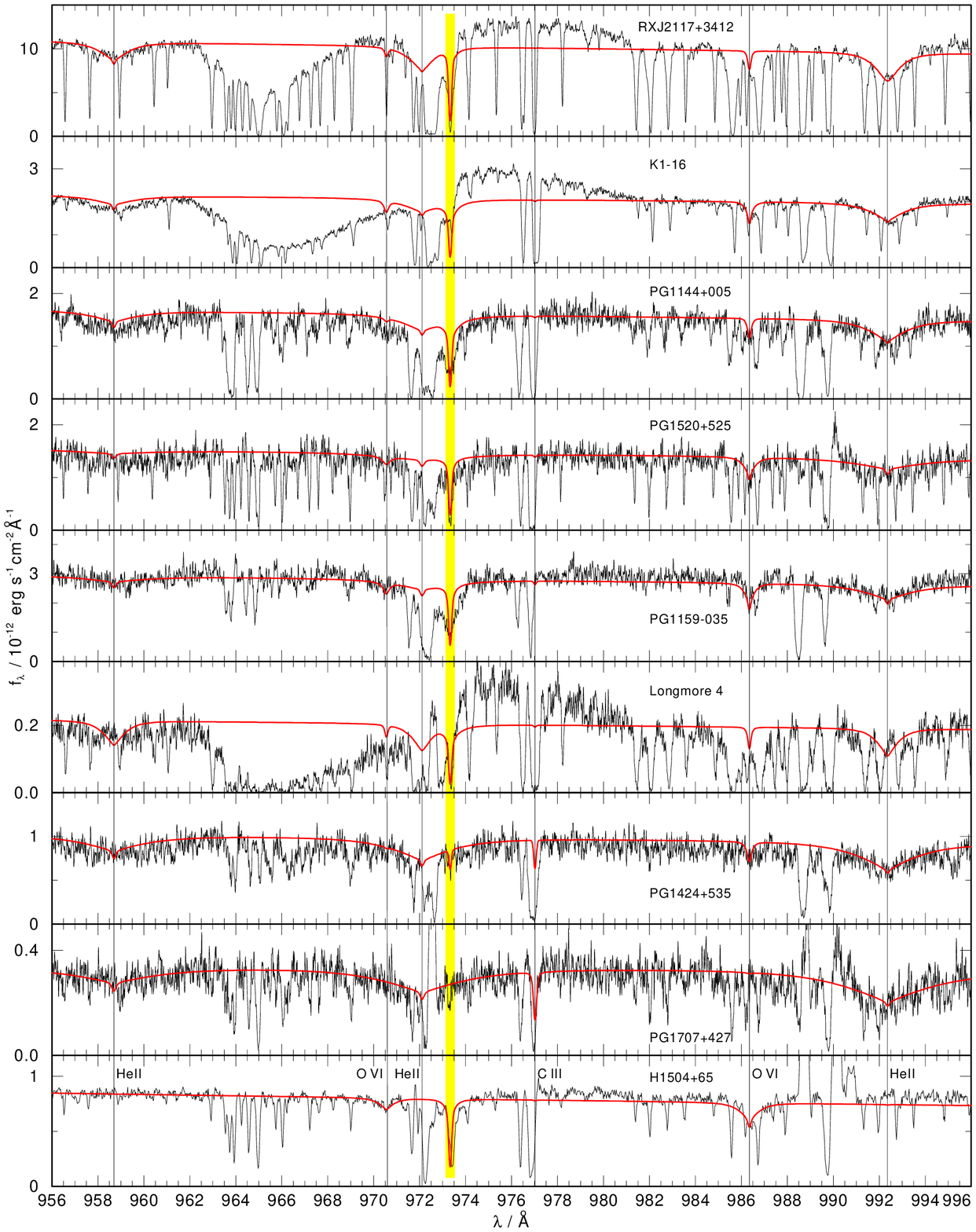}
  \caption[]{
Detail from FUSE spectra of PG1159 stars showing the vicinity of the newly
        identified \ion{Ne}{vii} line at 973.3\AA\ (indicated by the vertical
        shaded bar). \rxj, \keins, and Longmore~4 display broad and shallow
        P~Cyg profiles from the \ion{C}{iii}~977\AA\ resonance line. All other
        features, except for the labeled \ion{He}{ii} and
        \ion{O}{vi} lines, are of interstellar origin. The narrow emission
        features are terrestrial airglow. The spectra are smoothed
        by Gaussians with 0.02\AA\ FWHM.  Overplotted are model atmosphere
        spectra (Ne=2\%), folded with 0.1\AA\ FWHM Gaussians to match the
        instrumental resolution.  }
  \label{fuse_ne7_overview}
\end{figure*}

\begin{figure*}[tbp]
\includegraphics[width=17.8cm]{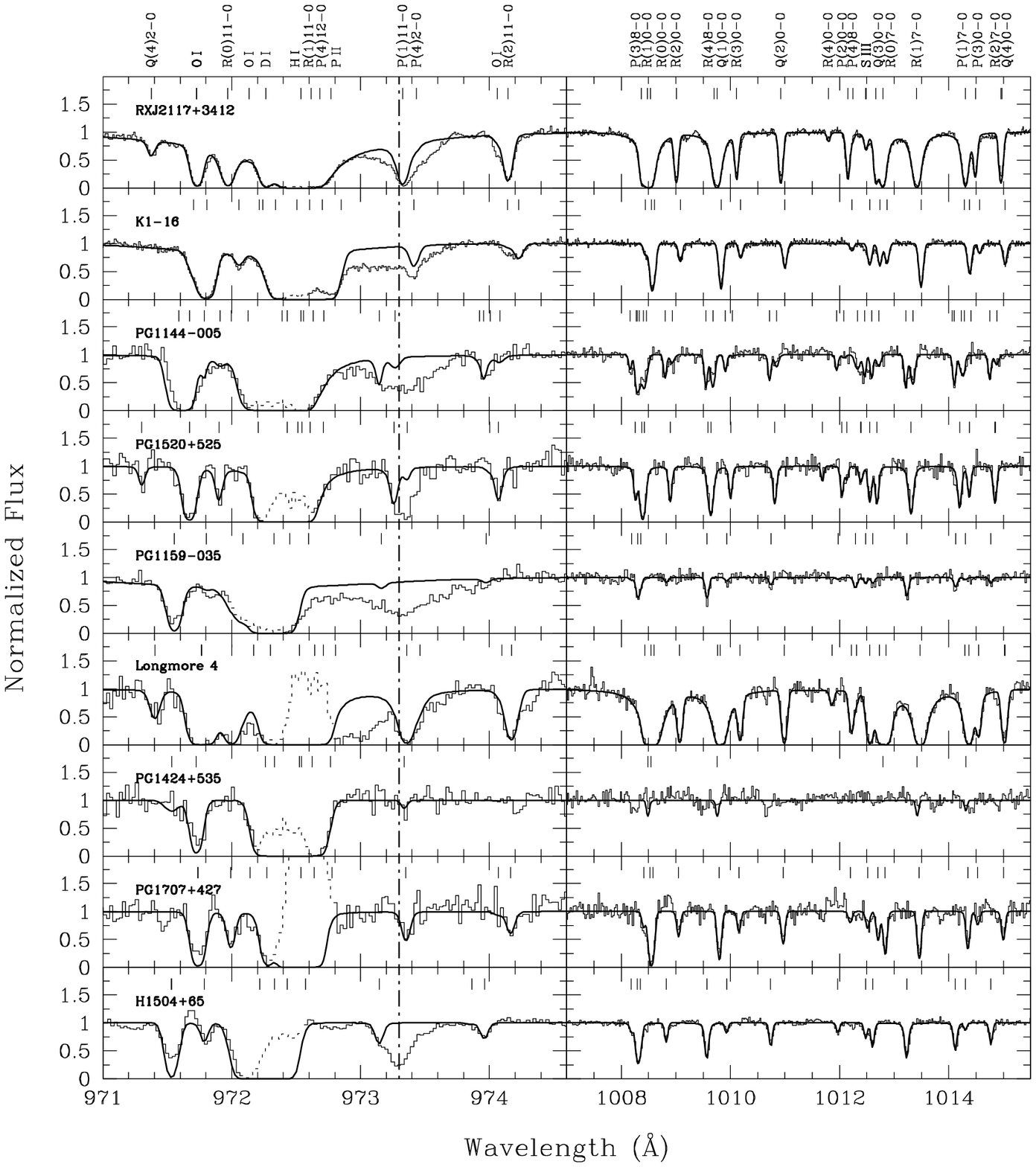}
  \caption[]{
Detail from FUSE spectra of PG1159 stars showing the modeled ISM absorption.
	The data are plotted as the thin histogram and the ISM model is the
        smooth heavy line.  
	The spectra were binned by 2 pixels in the top 2
        panels and by 4 pixels ($\sim$0.027\AA) in the others for plotting 
        purposes.  The panels on the left show the region in the vicinity of
        the newly-identified \ion{Ne}{vii} line at 973.3\AA\ (indicated by 
        the vertical dot-dashed line); the panels on the right show a 
        representative sample of \Htwo\ absorption.  The ISM features are
        labeled at the top of the figure and marked in each plot with 
        tickmarks.  The spectra were shifted to place the photospheric
        \ion{O}{vi} absorption at rest, so the ISM features appear with
        varying velocity offsets.  Airglow Ly $\gamma$ emission is plotted
        with a dotted line; in some spectra \ion{O}{i} 971.738\AA\ airglow 
        emission partially fills-in the corresponding IS absorption.
        }
  \label{fuse_ne7_ism}
\end{figure*}

\begin{figure}[tbp]
\includegraphics[width=8.1cm]{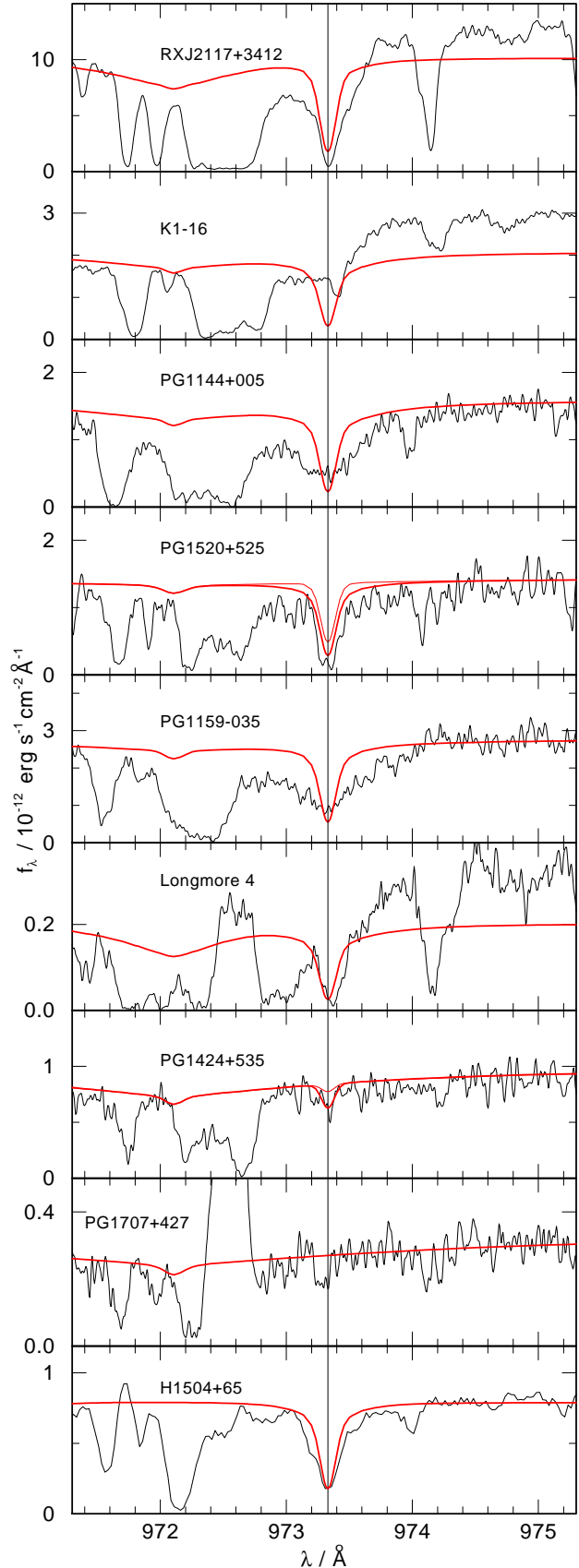}
  \caption[]{
Expanded view of Fig.\,\ref{fuse_ne7_overview}, centered around the
            \ion{Ne}{vii}~973.3\AA\ line. For PG1520+525 and PG1424+535 we
            overplotted a second synthetic line profile (thin line), computed
            with a solar Ne abundance (0.1\%).  }
  \label{fuse_ne7_detail}
\end{figure}

Fig.\,\ref{fuse_ne7_overview} displays the FUV spectra around the newly
identified \ion{Ne}{vii} line; an expanded view is shown in Figures 
\ref{fuse_ne7_ism} and \ref{fuse_ne7_detail}. 
They have been shifted to the rest wavelengths of
the photospheric lines, using the \ion{O}{vi} resonance line doublet.  We have
excluded available FUSE spectra of NGC\,7094 and Abell~78 from our analysis,
because the region around the \ion{Ne}{vii} line is badly contaminated by the
extremely broad P\,Cygni-shaped profile of the \ion{C}{iii} resonance line as
well as by a damped \Htwo\ profile. Their analysis requires detailed wind
modeling which is beyond the scope of this paper. The FUSE spectrum of another
object (HS\,2324+397) has been excluded from our analysis because of poor S/N
and very strong contamination by interstellar H$_2$ absorption lines.

The \ion{Ne}{vii} 973\AA\ line is potentially blended with two interstellar
\Htwo\ lines:  the Lyman band P(1)$_{11-0}$ 973.342\AA\ and the  Werner band
P(4)$_{2-0}$ 973.452\AA\ transitions.  The strengths and relative velocities of
these features were determined from fits to the numerous other \Htwo\ absorption
lines in the FUSE bandpass, along with the common low-ionization atomic species
(\ion{O}{i}, \ion{N}{i}, \ion{Ar}{i}, \ion{C}{ii}, \ion{Fe}{ii}, etc.).  The
SiC2a, LiF1a, and LiF2a spectra were used for the fits, as they had the best
signal-to-noise.  Fitting was performed with the program Owens.f written by
Martin Lemoine; for a discussion of the fitting procedure see Kruk \etal (2002)
or H\'{e}brard \etal (2002).  Atomic data used in the fits were taken from
Morton (2003), and molecular data were taken from Abgrall \etal (1993a, 1993b).
The P(4)$_{2-0}$ transition is rarely significant: the column density of \Htwo\
in the J=4 state is typically far less than in J=1, and absorption by \Htwo\
from the J=4 level was detected only for four of the nine stars.  Because
of the complications associated with modeling the broad wind features in
several of the stars, the stellar continuum was treated in the ISM models by
fitting a fourth-order polynomial over a wide region and including broad
\ion{He}{ii} profiles when appropriate.  The results of the fits are shown in
Fig.\,\ref{fuse_ne7_ism} for  the vicinity of the \ion{Ne}{vii} line, and for a
region spanning a  representative sample of \Htwo\ lines.  This latter region
includes the strongest Lyman band transitions and absorption from all J-levels
found in the spectra.  Fits of comparable quality are obtained throughout the
FUSE bandpass, so the modeled \Htwo\ absorption in the vicinity of 973.3\AA\
should be  reliable.  Results for individual lines of sight are discussed in
Section~\ref{sect_indiv} below.

The optical spectra presented here were taken between July 2000 and August 2002
in the course of the ESO Supernovae Ia Progenitor Survey (SPY), see
Tab.\,\ref{spectra_tab} (VLT/1 column).  SPY is a high-resolution spectroscopic
survey for radial velocity variable white dwarfs (Napiwotzki \etal 2001,
2003).  Spectra were taken with the high resolution UV-Visual Echelle
Spectrograph (UVES) of the UT2 telescope (Kueyen) of the ESO VLT. The SPY
instrument setup (Dichroic~1, central wavelengths 3900\AA\ and 5640\AA) used
UVES in a dichroic mode and nearly complete spectral coverage from 3200\AA\ to
6650\AA\ with only two $\approx$80\AA\ wide gaps at 4580\AA\ and 5640\AA\ is
achieved. SPY was implemented as a service mode program, which took advantage of
those observing conditions, which are not usable by most other programs (moon,
bad seeing, clouds).  A wide slit ($2.1''$) is used to minimize slit losses and
a $2\times 2$ binning is applied to the CCDs to reduce read-out noise. Our wide
slit reduces the spectral resolution to $R=18\,500$ (0.2\AA\ at 3600\AA) or
better, if seeing disks were smaller than the slit width.

\begin{figure*}[tbp]
\includegraphics[width=\hsize]{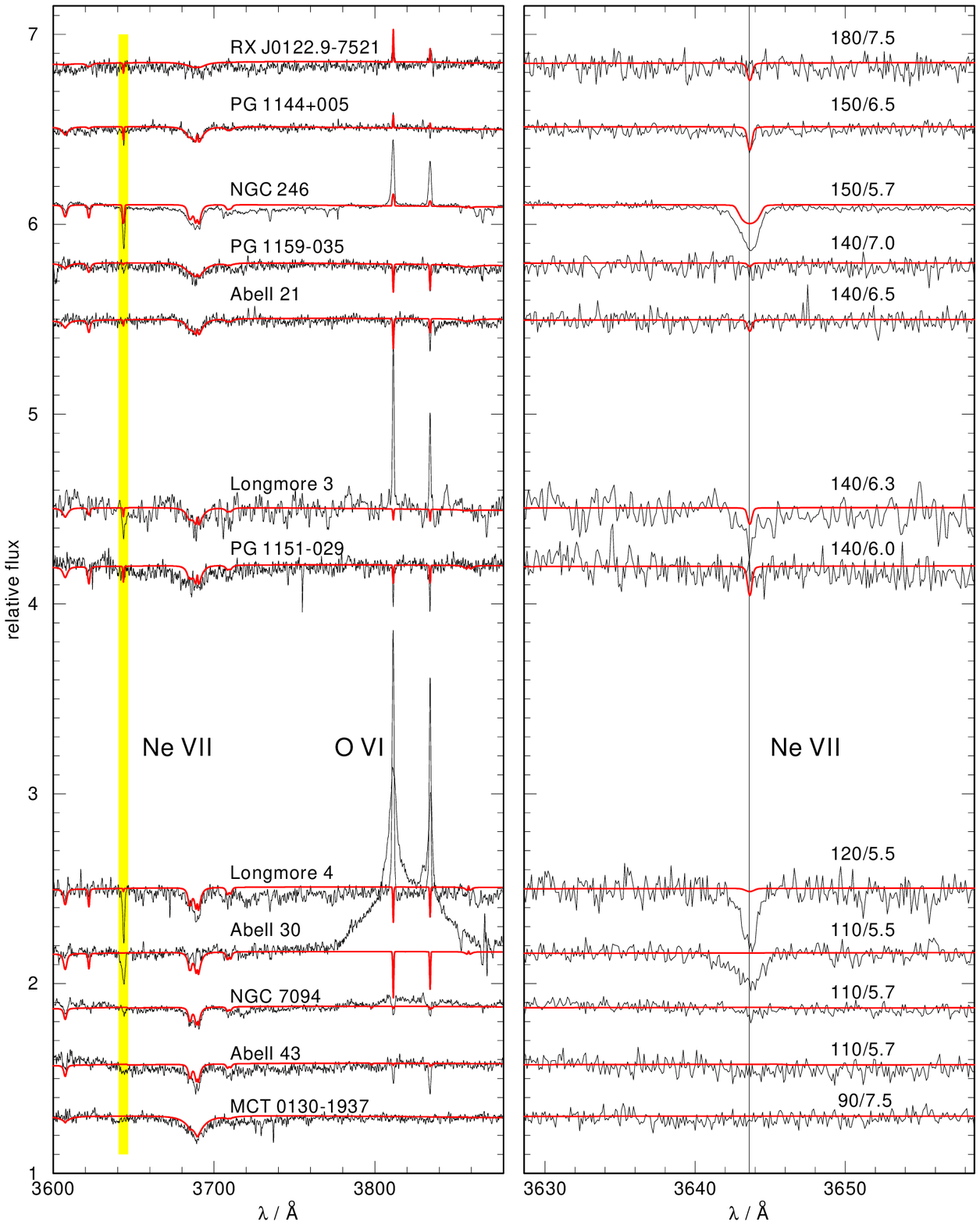}
  \caption[]{
Left: VLT spectra of PG1159 stars in the vicinity of the \ion{Ne}{vii}~3644\AA\
            line(indicated by the vertical shaded bar) and the
            \ion{O}{vi}~3s--3p doublet at 3811/3834\AA. Right: Expanded view
            centered around the \ion{Ne}{vii} line. Overplotted are model
            spectra with Ne=2\%, labeled with their \Teff\ (in kK) and
            \logg. Smoothing of observed spectra with Gaussians (FWHM given):
            left panel: 0.4--0.7\AA, right panel: no smoothing, except for
            Longmore 3 (0.2\AA). The model spectra are smoothed accordingly. The
            NGC\,246 model is also rotationally broadened (70\,km/s).  }
  \label{ne7_3644}
\end{figure*}

The spectra were reduced with a procedure developed by Karl (in prep.)  using
the ESO MIDAS software package, partly based on routines from the UVES pipeline
reduction software provided by ESO (Ballester \etal 2003).  Since the sampling
of the spectra is much higher than needed for our analysis, we 
rebinned the spectra to 0.1\AA\ stepsize; this considerably improved the 
signal-to-noise ratio with a negligible degradation of the resolution.
Figures~\,\ref{ne7_3644} and \ref{vlt_new_ne7_multiplet}
display details of the spectra near the \ion{Ne}{vii}~3644\AA\ and
3850--3910\AA\ lines, respectively.

\begin{figure}[tbp]
\includegraphics[width=\hsize]{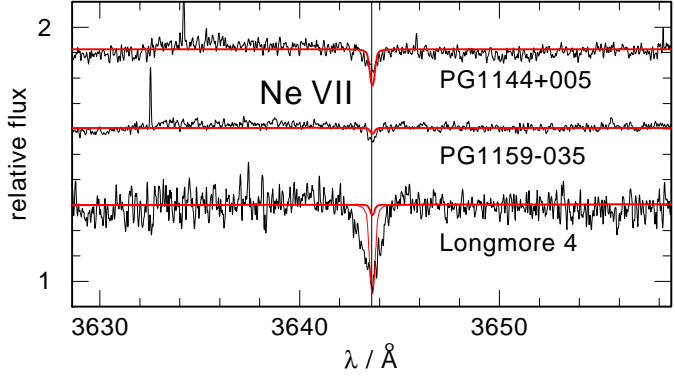}
  \caption[]{
Second set of VLT spectra of three PG1159 stars in the vicinity of the
            \ion{Ne}{vii}~3644\AA\ line, taken with higher spectral resolution. In
            comparison with the first set of spectra from these stars
            (Fig.\,\ref{ne7_3644}), we can now clearly identify the
            \ion{Ne}{vii} line in \elf.  Overplotted are the same model spectra
            as in Fig.\,\ref{ne7_3644}. A second model is overplotted with
            Longmore~4, namely the NGC~246 model from Fig.\,\ref{ne7_3644} (but
            unrotated). Smoothing of observed and computed spectra is performed
            with 0.05\AA\ and 0.07\AA\ Gaussians, respectively.  }
  \label{ne7_3644_deep}
\end{figure}

As part of another observation program, VLT/UVES spectra with significantly
higher spectral resolution and integration time  have been taken of \elf, \elv,
and Longmore~4 between April and July 2002 (see Tab.\,\ref{spectra_tab}, VLT/2
column).  We used the standard DIC1 (346) setup and a slit width of $0.6''$  and
arrive at a resolving power $R\approx 75\,000$ (0.05\AA\ at 3600\AA).  The
spectra were subject to the standard pipeline reduction provided by ESO. They
cover the wavelength regions 3300--3870\AA, 4800--5755\AA, and 5835--6805\AA.
Figure~\ref{ne7_3644_deep} displays these spectra near the
\ion{Ne}{vii}~3644\AA\ line.

\begin{figure}[tbp]
\includegraphics[width=\columnwidth]{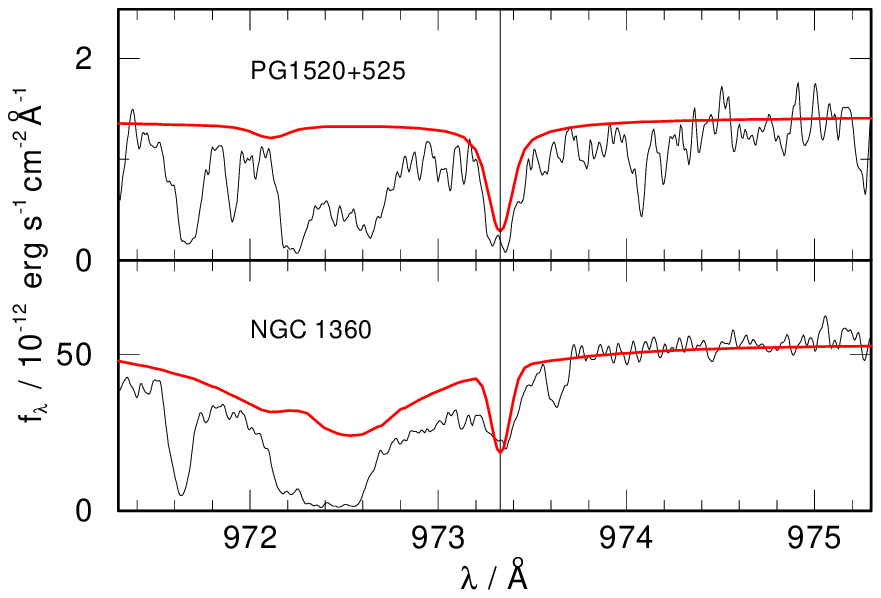}
  \caption[]{
Comparison of a PG1159 star (PG1520+525) and a ``normal'' hot hydrogen-rich
            central star (NGC\,1360). The \ion{Ne}{vii}~973.3\AA\ line is also
            detectable in NGC\,1360, which has a solar, i.e.\ much lower
            neon abundance.
            }
  \label{ngc1360}
\end{figure}

\section{Model atmosphere calculations}

For each program star we calculated a single model atmosphere with parameters
obtained from literature (Table\,\ref{objects_tab}). 
It is beyond the scope of this paper to improve these parameters by
detailed model grid calculations and line profile fits for the complete
wavelength range covered by the FUSE and VLT spectra. This is deferred to an
extensive analysis which has only just begun.

Line blanketed NLTE model atmospheres were computed using our {\sc PRO2} code
(Werner \etal 2003b).
The models are in hydrostatic and radiative equilibrium, hence, they cannot
reproduce the P\,Cygni shaped profile of the \ion{C}{iii} resonance line
observed in three objects. Detailed wind modeling will be necessary in future
analyses. The atomic models account for the most abundant elements in PG1159
stars, namely, helium, carbon, oxygen, and neon.  Mass fractions were as
shown in Table\,\ref{objects_tab}, except that the neon abundance was set to
2\%. We do not take into account the trace
nitrogen discovered in some of our program stars. This does not affect our
results obtained here. For more details on atomic models, we refer to {\small
W04}.

Here we present the detection of the \ion{Ne}{vii}~973.3\AA\ absorption line
2p$^1$P$^{\rm o}$--2p$^2$\,$^1$D in several PG1159 stars. The search was
initiated by the very first detection of this line in \hh\ ({\small W04}).
There is some uncertainty in the literature concerning the accurate wavelength
position of this line. The first measurement was published by Johnston \& Kunze
(1971, {\small JK69}) who give 973.6\AA. Lang (1983) quotes 973.33\AA\ (without
reference) and this value is also found in the Chianti database (Young \etal
2003). According to Kramida (NIST, priv.\ comm.)  the best measurement for this
line was done by Lindeberg (1972). The measured wavelength is
973.302$\pm$0.005\AA. For our synthetic spectra we have adopted 973.33\AA.

\begin{table}
\caption{\ion{Ne}{vii} lines identified in UV (vacuum wavelengths) and optical
  (air wavelength) spectra of PG1159
  stars. The last column gives the excitation energy of the lower level of the
  line transition. Level energies for the 3853--3912\AA\ multiplet are from
  Kelly (1987), but line positions determined from our observations. As a
  consequence, the uncertainty in Kelly's energy for the lower level 
  is x$\lppr$1600\,cm$^{-1}$.
\label{ne7_lines_tab}
}
\begin{tabular}{rllr}
      \hline
      \hline
      \noalign{\smallskip}
$\lambda$/\AA &  Transition & reference & E/cm$^{-1}$\\
      \noalign{\smallskip}
     \hline
      \noalign{\smallskip}
973.3  & 2p$^1$P$^{\rm o}$ -- 2p$^2$\,$^1$D     & Lindeberg (1972) & 214\,952\\
      \noalign{\smallskip}
\hline
      \noalign{\smallskip}
1982.0  \\
1992.1 & 3s$^3$S -- 3p$^3$P$^{\rm o}$ & Wiese \etal (1966)& 978\,320\\
1997.3  \\
      \noalign{\smallskip}
\hline
      \noalign{\smallskip}
3643.6 & 3s$^1$S -- 3p$^1$P$^{\rm o}$ & {\small JK69} &998\,250\\
           \noalign{\smallskip}
\hline
      \noalign{\smallskip}
3853.3&                              &            &1\,028\,386+x\\
3866.8&                              &            &1\,028\,519+x\\
3873.2&                              &            &1\,028\,519+x\\
3894.0& 3p$^3$P$^{\rm o}$ -- 3d$^3$D & this paper &1\,028\,775+x\\
3905.1&                              &            &1\,028\,775+x\\
3912.3&                              &            &1\,028\,775+x\\
           \noalign{\smallskip}
\hline
     \end{tabular}
\end{table}

In Table\,\ref{ne7_lines_tab} we list the UV/optical \ion{Ne}{vii} lines
detected in PG1159 stars. Besides the lines discussed here, we include for
completeness the \ion{Ne}{vii} triplet around 1992\,\AA\ which was identified in
one object (NGC\,246, {\small WR94}).

Our model calculations show that the 973\AA\ line of \ion{Ne}{vii} is always
much stronger than the optical 3634\AA\ line. This is because the excitation
energies of the (lower) levels involved are very different (see
Table\,\ref{ne7_lines_tab}). The level excitation energy of the 973\AA\ line
corresponds to an excitation temperature of 310\,kK, while that of the 3634\AA\
line corresponds to 1400\,kK. As a consequence, the population density of the
lower level of the FUV line is by many orders of magnitude larger than that of
the optical line (at least by four orders, depending on the model parameters and
the depth in the atmosphere). Therefore in some objects the FUV line can be
prominent while the optical line is not detectable.  It is obvious that the
\ion{Ne}{vii} 973\AA\ line is very strong in many of our program stars. In these
cases, the depth of the 973\AA\ line core is similar to that of the \ion{O}{vi}
resonance line, and both form in the uppermost layers of the atmospheres.  The
\ion{Ne}{vii} 973\AA\ line is quite saturated so that it is not very strongly
dependent on the neon abundance. This can be seen in Fig.\,\ref{fuse_ne7_detail}
for the case of PG1525+525. In the respective panel we included a model profile
with a Ne abundance reduced to the solar value, i.e. by a factor of 20 down to
0.1\%. The equivalent width is reduced by only a factor of about two. In
contrast, the optical 3634\AA\ line reacts sensitively on abundance changes. In
all of our program stars in which we see this optical line, it would be
undetectable if the neon abundance were only solar.

On the other hand, the  \ion{Ne}{vii} 973\AA\ line opens the possibility to
detect neon in very hot stars even at a solar neon abundance level. In
Fig.\,\ref{ngc1360} we show the FUSE spectrum of the very hot (\Teff=110\,kK,
\logg=5.6; Hoare \etal 1996) hydrogen-{\it rich} central star of NGC\,1360 which
has (at least roughly) a solar abundance composition. The \ion{Ne}{vii} line is
clearly visible and it is well matched by a model calculated with solar element
abundances.

\begin{figure*}[tbp]
\includegraphics[width=\textwidth]{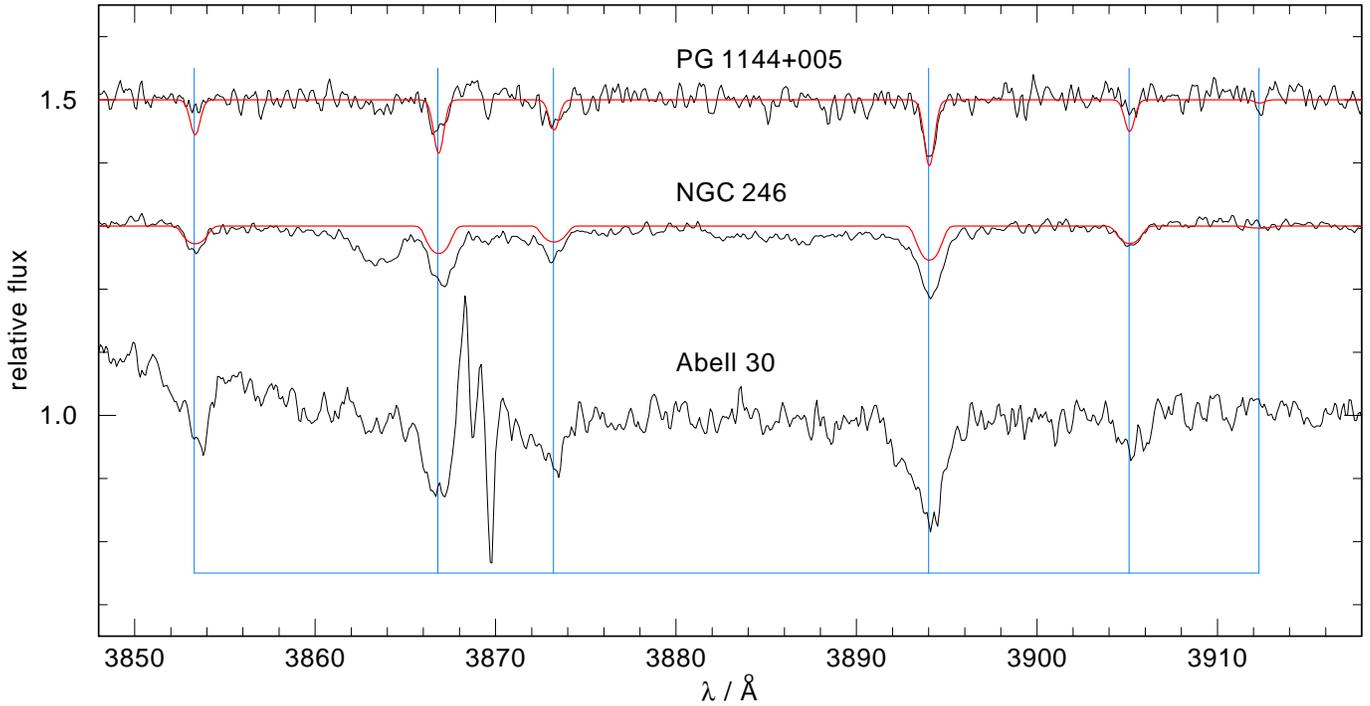}
  \caption[]{Identification of the \ion{Ne}{vii} 3p$^3$P$^{\rm o}$--3d$^3$D multiplet in three
    PG1159 stars. The vertical bars indicate the observed line positions
           which lie about 6\AA\ redward of the expected position when using Kelly's
           energy levels. Overplotted are the models for \elv\ and \ngc. The emission/absorption features between
           3869--3870\AA\ in Abell~30 are residuals from the subtraction
           of a nebular emission from \ion{[Ne}{iii]}~3869.06\AA. 
           }
  \label{vlt_new_ne7_multiplet}
\end{figure*}

A new \ion{Ne}{vii} multiplet, consisting of six lines in the 3853--3912\AA\
range, is identified in our VLT spectra of three PG1159 stars
(Fig.\,\ref{vlt_new_ne7_multiplet}), which means the first detection of this
multiplet in any stellar spectrum. It is the 3p$^3$P$^{\rm o}$--3d$^3$D
transition, whose exact line positions were unknown up to now because of
uncertain level energies. We have taken the energy levels from Kelly
(1987). If we shift the whole corresponding multiplet by $-6$\AA\ then all
predicted line positions fall into observed absorption features within $\pm
1$\AA. This enables us to determine the line positions to an accuracy of about
0.2\AA\ (vertical bars in Fig.\,\ref{vlt_new_ne7_multiplet}). This is a nice
example how high-resolution stellar spectroscopy can improve our knowledge of
atomic data.

As a side note, it is interesting to recall that Heap (1975) mentioned the
presence of a line feature at 3867.5\AA\ in her spectrum of NGC\,246. Now, after
three decades, it is evident that she saw for the first time  the 3866.8\AA\
line of that \ion{Ne}{vii} multiplet.

\section{Comparison of observed and theoretical neon line profiles}

Let us discuss the strengths of the \ion{Ne}{vii} lines in the FUSE and VLT
spectra and compare them to our models. As already mentioned, all the synthetic
line profiles were computed with a neon abundance of 2\%.  We begin with the
objects whose FUV spectra are displayed in Fig.\,\ref{fuse_ne7_overview} (and in
more detail in Fig.\,\ref{fuse_ne7_detail}), and then we turn to the objects for
which we have only the optical VLT data (Figs.\,\ref{ne7_3644},
\ref{ne7_3644_deep} and \ref{vlt_new_ne7_multiplet}).

As already emphasized, the 973\AA\ line is potentially contaminated by
interstellar \Htwo. It is therefore mandatory to carefully examine its possible
contribution to the photospheric neon line. We will see that the 973\AA\ line in
most objects appears significantly broader than predicted by our
models. Possible origins are discussed in Section~\ref{sect_broad}.

\subsection{Discussion of individual objects}\label{sect_indiv}

\rxj: The 973\AA\ line is located in the broad and shallow emission part of the
\ion{C}{iii} P\,Cygni profile. The heavily-saturated interstellar \Htwo\
P(1)$_{11-0}$ line coincides with the predicted \ion{Ne}{vii} profile, rendering
it difficult to confirm or refute its presence.  There is substantial residual
absorption, however, that is much broader than predicted by the photospheric
model, and which cannot be explained by interstellar absorption.

\keins: The weak narrow absorption seen at 973.4\AA\ is consistent with 
absorption by IS \Htwo.  There is substantial residual absorption, however, 
that is even broader than that seen in \rxj. The 3644\AA\ line was already 
detected by {\small WR94}.

\elv: The 973\AA\ line profile is affected by the red wing of the interstellar
\ion{H}{i} Ly$\gamma$ absorption line. The IS \Htwo\ absorption is quite weak
and offset significantly in wavelength. The neon  line is very strong and
significantly broader than the computed profile. Increasing \logg\ within
reasonable limits does not improve the fit, see also the discussion of the next
object.  We also identify the 3644\AA\ line. It is rather weak and well fit by
the synthetic profile. The strongest components of the
\ion{Ne}{vii}~3853--3912\AA\ multiplet are also identified and we obtain a good
model fit.

\fuenf: The 973\AA\ line is very strong and the synthetic profile fits very
well. The observed line profile is in strong contrast to the previous
object. Both stars have the same \Teff.  The surface gravity of \elv\ is {\it
lower} by one dex but, quite unexpectedly, its 973\AA\ line is much {\it
broader}.  The IS \Htwo\ contamination is a bit stronger than in \elv, but is
still a minor contributor to the overall profile.

\elf: Here we also see a strong 973\AA\ line, which is significantly broader
than what we expect from the synthetic profile.  The IS \Htwo\ contamination is
not significant.  This star's gravity is also {\it lower} than that of \fuenf,
but the line profile is much {\it broader}.   The 3644\AA\ line is not
detectable in the VLT/1 spectrum of \elf\ (Fig.\,\ref{ne7_3644}), but easily
identified, although being quite weak, in the higher-resolution VLT/2 spectrum
(Fig.\,\ref{ne7_3644_deep}). Accordingly, the model profile is very weak,
too. The \ion{Ne}{vii}~3853--3912\AA\ is detectable (not shown).

Longmore~4: As with \rxj, the heavily-saturated IS \Htwo\ absorption coincides
with and obscures the predicted \ion{Ne}{vii} absorption.  There is, however,
very strong unexplained absorption that is blue-shifted from the predicted line
center. Among all the VLT spectra, the 3644\AA\ line of Longmore~4 is the
strongest (together with NGC\,246). However, the computed profile for Longmore~4
is much too weak. We believe that this is partly due to the fact that its
temperature is significantly higher than previously determined. The observed
strong \ion{O}{vi}~3811/3834\AA\ doublet is also in contrast to the weak
computed emission and, hence, points into the same direction.  Instead of
120\,kK, \Teff\ could be as high as that of NGC\,246 (150\,kK). In the highest
resolution spectrum (Fig.\,\ref{ne7_3644_deep}), the optical \ion{Ne}{vii} line
appears asymmetric, showing a blue wing that is broader than the red one. We
think that this is due to wind effects. So the line equivalent width in this
star is stronger than predicted also because of our static model assumption.

\vier\ and \sieben\ are among the ``cool'' PG1159 stars and our models predict
rather weak 973\AA\ lines, because the ionization balance of neon shifts
strongly in favor of lower ionization stages.  \Htwo\ is weak but clearly
detected in \sieben, and appears sufficient to explain the observed absorption
at 973\AA.  \Htwo\ is not convincingly detected in the spectrum of \vier;
instead what is shown in  Fig.\,\ref{fuse_ne7_ism} is a 3-$\sigma$ upper limit
to the \Htwo\ column density.  The resulting profile is consistent with the
observed spectrum.  The signal-to-noise of the FUSE spectra for both \vier\ and
\sieben\ is relatively low; better data are required to determine conclusively
whether or not the predicted weak \ion{Ne}{vii} absorption is present.

We have augmented Figs.\,\ref{fuse_ne7_overview} -- \ref{fuse_ne7_detail}
with the FUSE spectrum of \hh, which is the hottest PG1159 star (200kK). It
shows strong \ion{Ne}{vii} lines, both, in the FUV and the optical ({\small
W04}).  

We now turn to those objects for which only optical data are available
(see Fig.\,\ref{ne7_3644}). We remark in advance, that the \ion{O}{vi}~3s--3p
doublet at 3811/3834\AA\ may be strongly enhanced in emission strength by a
stellar wind, so that we cannot expect good fits of our models to these
lines. This is most obvious in the case of Abell~30.

\rxja\ is the second hottest PG1159 star (\Teff=180\,kK) after \hh\
(200\,kK). The 3644\AA\ line cannot be detected. Also, the predicted profile is
very weak and shows that the line can escape detection despite the
overabundance of Ne. When compared to \hh\ this seems surprising. But the
surface gravity of \rxja\ is lower, favoring the shift of the neon ionization
balance to ionization stages higher than \ion{Ne}{vii}.

NGC\,246 has the strongest 3644\AA\ line, together with Longmore~4. Its
equivalent width is significantly stronger than the computed one.  Five out of
the six components of the \ion{Ne}{vii}~3853--3912\AA\ multiplet are identified
and, again, the observed equivalent widths are larger. This suggests that the Ne
abundance in both stars may be even higher than 2\%. We recall that a similar
result was found for \hh\ (see Sect.\,1). Note that NGC\,246 is rotating,
perhaps with up to v\,sini$\approx$70km/s (Rauch \& Werner 1997), which can
partly explain the broad line profile.

Abell~21: The optical lines cannot be detected. The synthetic profiles show
that, as in the case of \elf, the relatively high surface gravity shifts the
ionization balance away from \ion{Ne}{vii} to \ion{Ne}{vi}.

Longmore~3: The observed 3644\AA\ line is much stronger than the calculated
one. In this case we think that either \Teff\ (140\,kK) is higher than assumed
and/or that \logg\ (6.3) is lower. Both would increase the calculated line
strength. This is corroborated by the appearance of the
\ion{O}{vi}~3811/3834\AA\ doublet: The calculated lines are in absorption
whereas the observed ones are in emission. This also points to higher \Teff\
and/or lower \logg.

PG1151-029: The optical lines cannot be detected. The computed profiles show
that they can escape detection in our VLT spectrum.

Abell~30 presents a strong 3644\AA\ line but nothing is seen in the model. As
in the case of Longmore~4, \Teff\ is probably underestimated and/or \logg\ is
overestimated. Out of all objects presented here, the
\ion{Ne}{vii}~3853--3912\AA\ multiplet is most prominent in Abell~30. Again, the
lines do not show up in the model, because its parameters are not
appropriate. 

NGC\,7094 and Abell\,43 are the two hybrid-PG1159 stars in our sample.  They are
relatively cool so that only a weak 3644\AA\ line can be expected. Indeed, a
very weak line is possibly present in NGC\,7094, but we regard this as
uncertain. It is not detectable in Abell~43.

\mct\ is the coolest object in our VLT sample (90\,kK). Consequently, the
optical lines are not detectable.

\subsection{Summary of neon line analyses}\label{sect_broad}

Table~\ref{results_tab} summarizes the results. \ion{Ne}{vii} lines were not
detected in seven objects: \rxja, Abell~21, PG1151-029, \vier, Abell~43, \mct,
and \sieben. In all these cases one can neither exclude nor prove a neon
overabundance, because the respective atomic levels are weakly populated
due to photospheric parameters.  In the other ten objects, neon lines are
detected.  A comparison with model profiles reveals a strong overabundance
  in some of these, however, a detailed abundance analysis still needs to be
  performed for all of them.

In six PG1159 stars we have detected the \ion{Ne}{vii}~973\AA\ line. It is
remarkable that the observed line profile in five of these stars is much broader
than the synthetic profile (see Tab.\,\ref{results_tab}). Can wind effects,
neglected in our models, play a role? Probably not, at least not in all
cases. While three of these five stars show P~Cyg profiles, two don't: \elf\ and
\elv. Another explanation could be line broadening by pulsations. Four out of
five of these broad-line PG1159 stars are non-radial pulsators (identified in
Tab.\,\ref{results_tab}). \elf\ is indeed the prototype of the GW~Vir pulsators,
and periodic UV line profile shifts with an amplitude of the order 5\,km/s  were
detected in HST/STIS spectra (Dreizler \& Werner 2004). The \ion{Ne}{vii} line 
in the FUSE spectrum would require a much larger
amplitude. This is not necessarily a contradiction, because the weak HST-UV
lines (mostly \ion{C}{iv} and \ion{O}{vi} lines) form in deeper regions of the
atmosphere than the strong \ion{Ne}{vii}~973\AA\ line, and different depths may
experience different pulsation velocities.  Hence, pulsation could explain the
broad line profiles in \elf\ and the other broad-line PG1159 stars except for
one object, namely \elv. Several observation campaigns have failed to detect
light variability in this star (see, e.g., Steininger \etal 2003).

It is worth mentioning that the unusual width of the \ion{Ne}{vii}~973\AA\ line
in \elf\ and \elv\ comes along with a similar effect in their \ion{O}{vi}
resonance doublets. These doublets have optical depths comparable to the neon
line and also show broader and shallower cores than the models. In contrast, the
only star which has a narrow \ion{Ne}{vii}~973\AA\ line (that is compatible with
the model), also exhibits an \ion{O}{vi} line that is fitted by the model. An
interpretation of the oxygen doublet in the other four broad-line PG1159 stars
is problematic because it appears as a prominent P~Cyg profile.

\begin{table}
\caption{
Summary of \ion{Ne}{vii} line identifications in our program stars. ``+'' and
``-'' denote detection and non-detection, respectively. ``broad'' means that the
detected line is much broader than the computed one. No entry means that no
spectra are available. The second column denotes if the star is a pulsator (+)
or non-pulsator (-). No entry means, that no photometric observations were
performed or published.
\label{results_tab} }
\begin{tabular}{l c c c c}
\hline 
\hline 
\noalign{\smallskip}
Object & variable & 973\AA & 3644\AA & 3850-3910\AA \\
\noalign{\smallskip}
\hline
\noalign{\smallskip}
\noalign{\smallskip}
\rxja       & - &       & - & - \\  
\rxj        & + & broad &   &   \\
PG1520+525  & - &   +   &   &   \\
PG1144+005  & - & broad & + & + \\
NGC\,246    & + &       & + & + \\
PG1159-035  & + & broad & + & + \\
Abell 21    &   &       & - & - \\
\keins      & + & broad & + &   \\
Longmore 3  & - &       & + & - \\
PG1151-029  & - &       & - & - \\
Longmore 4  & + & broad & + & - \\ 
PG1424+535  & - &   -   &   &   \\
Abell 43    &   &       & - & - \\
NGC\,7094   &   &       & + & - \\
Abell 30    &   &       & + & + \\ 
MCT0130-1937& - &       & - & - \\ 
PG1707+427  & + &   -   &   &   \\
\noalign{\smallskip}
\hline
\end{tabular}
\end{table}

\section{Discussion and concluding remarks}

In order to discuss the implications of our results let us describe shortly our
current understanding about the origin of PG1159 stars. We refer the reader to
the work of Herwig (2001) for more details.

PG1159 stars are hot post-AGB stars which have suffered a late (or very late)
helium-shell flash during their first descent from the AGB. This event
transforms a post-AGB star (or a white dwarf) back to an AGB star
(``born-again'' AGB star) and it then descends from the AGB for a second
time. The He-shell flash is the origin of the H-deficiency that characterizes
the PG1159 stars (and their immediate progenitors, the [WC]-type central stars).

The surface abundance pattern of PG1159 stars essentially reflects the
composition of the matter that is located between the H- and He-burning shells
(intershell region) of thermally pulsing AGB stars.  Intershell matter is
helium-dominated (ash of CNO hydrogen-burning), but also strongly enriched by C
and O from the 3$\alpha$-process in the He-burning shell. The strong C and O
intershell enrichment is the consequence of effective convective overshoot that
dredges up C and O into the convective intershell. Evolutionary models predict a
neon abundance of about 2\% by mass in the intershell. This $^{22}$Ne is
produced in the He-burning shell from $^{14}$N (which itself was produced
previously by CNO-burning) via \\ \\ \mbox{}\hspace{2cm}
$^{14}$N$(\alpha,\gamma)^{18}$F$({\rm e}^+\nu)^{18}$O$(\alpha,\gamma)^{22}$Ne
. \\ \\  Our result, that the neon abundance in some objects is strongly 
overabundant, is therefore in qualitative agreement with theoretical predictions.

This view is further supported by recent observations which reveal s-process
signatures in the surface composition of PG1159 stars. While it is impossible to
detect the enrichment of s-process elements in these hot objects (because of the
lack of atomic data for high ionization stages), we have instead found a
deficiency of iron in all examined PG1159 stars (Miksa \etal 2002, Werner \etal
2003a). Obviously, Fe was transformed to heavier elements in the intershell
region of the AGB star  by n-captures from the neutron source
$^{13}$C($\alpha$,n)$^{16}$O.

How is the He/C/O/Ne-rich intershell matter mixed to the surface of the star,
and what happens to the H-rich envelope matter? This is interesting in order to
discuss the observed nitrogen abundance in PG1159 stars.

As a consequence of the He-shell flash a convection zone develops above the
He-burning shell that eventually reaches the H-rich envelope. Hydrogen is
ingested and burned. According to Iben \& MacDonald (1995) large amounts of
nitrogen can be produced (5\%) by the reaction chain \\ \\\mbox{}\hspace{2cm}
$^{12}$C(p,$\gamma$)$^{13}$N$({\rm e}^+\nu)^{13}$C(p,$\gamma$)$^{14}$N ,\\ \\
because ample $^{12}$C is available in the intershell from 3$\alpha$ burning.
However, calculations by Herwig (2001), with a more extensive nuclear network,
show that {\it no} fresh nitrogen is produced. Instead of $^{13}$C reacting with
protons to form $^{14}$N, it reacts with $^{4}$He to form $^{16}$O, so that
protons are burned in the chain \\ \\ \mbox{}\hspace{2cm}
$^{12}$C(p,$\gamma$)$^{13}$N$({\rm e}^+\nu)^{13}$C($\alpha$,n)$^{16}$O .\\ \\
Here we have an effective neutron  source which activates further n-capture
nucleosynthesis (Herwig \etal 2003).

If no fresh N is produced, and if all N from previous CNO cycling was
transformed into $^{22}$Ne, then the question remains unanswered, why a
considerable fraction of PG1159 stars (six objects) exhibits nitrogen on the 1\%
level. As a coincidence, these six N-rich stars are within our sample
(PG1144+005, PG1159-035, Abell~30, PG1707+427, Abell~43, NGC\,7094). The first
four of these positively show \ion{Ne}{vii} lines, i.e., the neon must have been
produced from CNO-cycle generated nitrogen.
 
The hybrid-PG1159 stars are a special case. We know four such objects which
still have considerable amounts of residual surface hydrogen (two are in our
sample: Abell~43 and  NGC\,7094). According to Herwig (2001) this is the result
of a He-shell flash experienced by an AGB-star immediately before its departure
from the AGB. They also show residual nitrogen, which seems natural in this
event. The hybrid-PG1159 stars  essentially display an intershell matter
abundance pattern like the other PG1159 stars, however, it is diluted by
hydrogen-rich envelope matter. So we also would expect a neon-enrichment, but
this is difficult to conclude from our spectra.

In summary, the strong neon enrichment found in some PG1159 stars
is qualitatively in agreement with evolutionary models for stars which suffered a
late He-shell flash. The surface chemistry of PG1159 stars is that of intershell
matter in the precursor AGB stars. The neon enrichment and the
iron-deficiency caused by s-processing are strongly supporting
this scenario.

\begin{acknowledgements}
UV data analysis in T\"ubingen is supported by the DLR under grant
50\,OR\,0201. RN acknowledges support by a PPARC advanced fellowship. 
JWK is supported by the FUSE project, funded by NASA
contract NAS5-32985. We thank Falk Herwig (LANL) for helpful discussions on
evolutionary aspects, and Dr.\ Kramida (NIST) and Prof.\ Kunze (University
of Bochum) for their advice on the \ion{Ne}{vii}~973\AA\ line
identification. This research has made use of the SIMBAD Astronomical Database,
operated at CDS, Strasbourg, France. The UVES spectra used in this analysis were
obtained as ESO Service Mode runs.  The interstellar absorption analysis was
done using the profile fitting procedure Owens.f developed by M.\ Lemoine and the
FUSE French Team.
\end{acknowledgements}

\end{document}